\documentstyle{aipproc}
\def\be{\begin{equation}}
\def\ee{\end{equation}}
\def\ltap{\ \raisebox{-.5ex}{\rlap{$\sim$}} \raisebox{.4ex}{$<$}\ }

\def\bsg{b\to sg}
\def\bsa{b\to s\gamma}
\def\bsqq{b\to sq\bar q}
\def\Ftg{F_2^g}
\def\Bsl{{\cal B}_{\rm s.l.}}
\def\nC{n_C}
\def\etap{\eta^\prime}

\def\Betap{B\to \eta^\prime + X_s}

\begin{document}
\title{New Physics and Enhanced Gluonic Penguin}

\author{George W.S. Hou}
\address{Department of Physics,
National Taiwan University,
Taipei, Taiwan 10764, R.O.C.}

%\lefthead{LEFT head}
%\righthead{RIGHT head}
\maketitle

\begin{abstract}
We discuss the historical development of the gluonic $B$-penguin,
its sensitivity to $H^+$ effects, and $\bsg\sim$ 10--15\% as 
a possible solution to the $\Bsl$ and $n_C$ problems.
The latter and the connection of the gluonic penguin to
inclusive $B\to \eta^\prime+ X_s$ production through the gluon anomaly, 
with the intriguing prospect of 10\% inclusive CP asymmetries,
bring us to topics of current interest.
\end{abstract}

\section{SM: Historical Backdrop}

We are here to celebrate the 20th anniversary of
the $\Upsilon$ discovery.
Because of this historic setting, 
I will dwell a little more on the historical aspects 
(from a personal perspective) of gluonic penguins,
before I turn to the current.

Shortly after 1977,
Bander, Silverman and Soni (BSS) \cite{BSS} suggested 
the mechanism of (direct) CP violation in the {\it decay} of $b$ quarks.
The  $B$-penguin was born, as illustrated in Fig. 1 (a). 
CP violation is possible because all 3 generations run in the loop,
while on-shell $u\bar u,\ c\bar c \to g^* \to q\bar q$ rescattering
provides the second $i$, an absorptive part.
Tony Sanda told me that this work was an inspiration for 
his mixing dependent CP violation ideas.

\begin{figure}[h]
\includegraphics{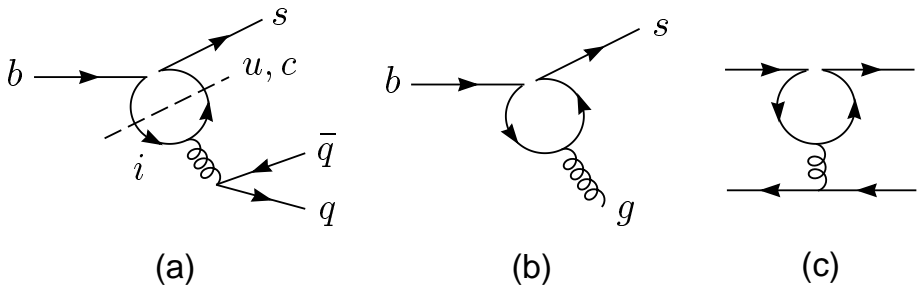}
\vskip 3.8cm
\caption{
(a) Timelike, (b) lightlike and (c) spacelike gluonic penguin.
The cut in (a) corresponds to on-shell $u\bar u$ and $c\bar c$ pairs, 
while in general $i = u,\ c,\ t$ in the loop.
}
\end{figure}

Perhaps influenced by BSS,
Guberina, Peccei and R\" uckl pointed out \cite{GPR} in late 1979,
using operator language, that  $B$-penguin operators 
``can alter the `natural' Cabibbo pattern of the Kobayashi--Maskawa model."
Namely, $B\to K\pi > B\to \pi\pi$ is possible, since 
the penguin is $\propto \vert V_{cb} \vert\simeq \vert V_{ts}\vert$, 
while tree level $B\to \pi\pi$ could be suppressed if $V_{ub}$ is very small.
This has just turned into fact this year with CLEO's observation of 
a few 2-body rare decays \cite{Poling}.

Following the study of $\bsa$ decay in the early 80's,
%which turned out to be extremely small,
Eilam \cite{Eilam} added $\bsg$ of Fig. 1(b)
to the ``parton" estimate of the inclusive penguin rate.
In 1987, the so-called ``large" QCD correction to $\bsa$ was discovered,
leading to the ``operator" approach industry.
But $\bsg$ changed only from 0.1\%
to 0.2\% with QCD corrections, much less dramatic than the $\bsa$ case.

The inclusive gluonic penguin was clarified \cite{HSS} in 1987
by noting the $q^2$ of $g^*$, i.e.
$\bsg^* \equiv \bsqq,\ sg,$ and $b\bar q\to s\bar q$
(timelike, lightlike and spacelike),
where the latter, given in Fig. 1(c),
is the familiar looking ``penguin" from kaon physics.
It was found that \cite{HSS} $\bsqq \sim 1\% \sim b\to u$ 
while $b\bar q\to s\bar q < \bsg$.
However, counter to one's intuition \cite{JYSW}, 
$b\to sgg \ll b\to sq\bar q$. 
Interestingly, the 3-body $\bsqq$ at ${\cal O}(\alpha_s^2)$
dominates over the 2-body $\bsg$ at ${\cal O}(\alpha_s)$,
which comes about because of a subtlety of GIM cancellation.

There are two conserved effective $bsg$ couplings;
ignoring $V_{ub}$ they are
\begin{equation}
G_{\rm F} g_s \; v_t \;
\bar s t^a
\{\Delta F_1\; (q^2\gamma_\mu - q_\mu \not{\! q}) L
- F_2\; i\sigma_{\mu\nu} q^\nu m_b R \} b,
\end{equation}
where $ \Delta F_1 \equiv F_1^t - F_1^c \simeq 0.25
- ( -2/3\, \log(m_c^2/M_b^2) -2/3\, \log(m_b^2/M_W^2)) \simeq -1.3 -2.75$,
and $F_2 \cong F_2^t \simeq 0.2$.
$F_1$ contains large logarithms while $F_2$ does not,
but suffers from power GIM suppression.
However, $F_1$ cannot contribute to $\bsg$ because of the $q^2$ factor.
Thus, the subtle higher order dominance comes about because of
having a logarithmic and a power GIM suppressed effective coupling,
and only the latter leads to $\bsg$.

\section{$H^+$ Effects: $F_2^\gamma$ and $\Ftg$}

The above subtlty leads to surprising $H^+$ effects:
$F_2^\gamma$ is very sensitive to low $m_{H^+}$. 
For the Higgs sector of minimal SUSY, 
the effect is always constructive 
and does not vanish with $\tan\beta$
(ratio of v.e.v.'s of the two Higgs doublets)  \cite{GW,HW},
which holds similarly for $F_2^g$ \cite{HW}.
Though $\bsg$ could not be greatly enhanced in this model, 
both strong enhancement/suppression of $\bsa$ are possible
for a second model, and $\bsg$ could become {\it very} enhanced \cite{HW}. 

At that time the experimental limit was ${\cal  B}(\bsa) < 6 \times 10^{-3}$, 
while $\bsg$ was practically without bound
(except $b\to c$ should be dominant).
The curious thing about $\bsg$ is that it does not lead
to any good, tangible signature!
By 1992, however, 
the CLEO limit on ${\cal  B}(\bsa)$ improved to $5\times 10^{-4}$,
entering the domain of SM predictions.
This had a dramatic implication that $m_{H^+} > 250$ GeV 
or so \cite{Hewett} in SUSY type models.
Unfortunately, 
because $bsg$ and $bs\gamma$ couplings in Higgs models
are highly correlated, the limit and eventual observation of 
$\bsa$ by CLEO meant that $\bsg$ could 
no longer be strongly enhanced in usual charged Higgs models \cite{GHT}.

\section{$\Bsl$--$\nC$ Problem: Enhanced $\bsg$}

Some indirect indications for enhanced $\bsg$ appeared, in fact,
in the early 90's.
As  $B$ experiments matured, the semileptonic branching ratio ($\Bsl$)
steadily declined, from $\simeq 12\%$ in 1986, to 10.7\% by early 1990.
Theory predicted 12--15\%, 
hence it appeared \cite{AP,GH} that 
the SM had trouble with the experimental value, 
with a 10--15\% discrepancy.
The relevant QCD scale $\mu$ for $B$ decay 
could be much lower than $m_b$ \cite{AP},
or one could have {\it new physics} $\Gamma_{\rm New} \sim $10--15\%,
which drives down $\Bsl$ via
\be
\Bsl = {\Gamma(B\to \ell\nu + X)/
       (\Gamma^{\rm SM}_{\rm tot.} + \Gamma_{\rm New})}.
\ee 
The new process must be relatively well hidden,
and low in charm content to accommodate the analogously
low charm counting rate \cite{Cassel} (the $n_C$ problem).
Two modes were suggested \cite{GH}, both from $H^+$  effects.
The first one, $B\to \tau\nu + X \sim 10\%$,
was very quickly %(I was startled by the shortness of shelf-life!)
ruled out by a superb analysis job of ALEPH, which confirmed SM predictions.
The second possibility of $b\to sg \sim $10--15\%,
which is a charmless final state, was very difficult to rule out.

However, by 1994, the possibility of enhanced $\bsg$ due to $H^+$ effects
became implausible because of $\bsa$ limits/measurements.
Subsequently, Kagan \cite{Kagan}
suggested that TeV scale physics responsible for quark mass (and mixing)
generation might lead to enhanced $\bsg$.
For example, gluonic insertions to $\bar s_{L}b_{R}$ mass terms
could result in effective $s_Lb_Rg$ couplings. 
To disentangle $\bsg$ from $\bsa$, 
one must employ more {\it color} in the loop.
Note that $H^+$ is colorless and does not couple to gluons, 
hence diagrams for $\bsg$ are only a subset of $\bsa$.
But gluons could more readily couple to gluinos via
SUSY $q_i \tilde q_j \tilde g$ couplings \cite{Kagan,CGG}
(with flavor violation in squark mass matrix),
or to techniscalars \cite{Kagan}.
In this way $\bsg$ could in principle be separated from $\bsa$ and
be strongly enhanced.

The problems of low $\Bsl$ and $n_C$ have persisted to this day, 
despite much theoretical and experimental effort.
Two recent analyses \cite{KR,DISY} give
$\Bsl = 0.105 \pm 0.005$ and $n_C = 1.10 \pm 0.06$,
both low by about 10--15\%.
The problem could still be experimental,
and in fact the latest results hint at softening of the problems,
but two views are offered on enhanced charmless $b$ decays.
Kagan and Rathsman \cite{KR} think that 
$\Bsl$, $n_C$ and kaon excess in $B$ decays 
together hint at $\bsg \sim $10--15\%. 
Using JETSET fragmentation of the $s$ quark, 
they find the $K$ spectrum to be rather soft, 
hence $\bsg$ indeed hides well.
On the other hand, Dunietz {\it et al.} \cite{DISY} suggest that 
half of $b\to sc\bar c$ (expected at 20--30\% level) 
has disappeared into light hadrons.
The effect has to be nonperturbative to evade
the perturbative $\bsg^* \sim$ 1\% discussed earlier.
The proposed mechanism is via a $c\bar c g$ hybrid meson,
since the $c\bar c$ pair is mainly in color octet configuration.
The hybrid should \cite{HT} be favorably produced in $b\to s c\bar c$,
should be relatively narrow (long lived)
and should have suppressed decays into $D\bar D+ X$ and usual charmonia.
Hence \cite{HT}, in a sense it is no less exotic than 
new physics $\bsg \sim$ 10--15\%.

\section{Inclusive $\etap$, $\bsg$, Gluon Anomaly}

1997 will be remembered as the year of the strong penguin.
Since the Aspen Winter Conference, 
CLEO has reported the first observations of a host of two-body 
rare $B$ decays.
$K\pi \sim 10^{-5}$ is observed, while $\pi\pi$ is not, 
confirming the GPR suggestion \cite{GPR}.
The $\omega h^\pm$ mode is larger than expected,
while $\eta^\prime K \sim 10^{-4}$ is huge, but $\eta K$ is not seen!
All in all, we see that penguins are large.

What is even more astounding is the observation of  \cite{Poling}
\begin{equation}
{\cal B}(B\to \eta^\prime + K + X) = (75 \pm 15 \pm 11) \times 10^{-5}
\mbox{\hskip 0.7cm \rm ($2.0 < p_{\eta^\prime} < 2.7$ GeV)}
\end{equation}
where $X = 0$--$4\, \pi \ (\leq 1\,\pi^0)$.
%and of order 80 events are seen.
While a cut on $p_{\eta^\prime}$ is in part to suppress background,
it is astonishing to see so many events in this
rather unusual channel.
%Who would have thought of large fraction of fast $\eta^\prime$'s beforehand?
If one extrapolates from Eq. (3),
one could easily saturate $\bsg^* \sim 1\%$.

The most prominent feature is that the $\etap$ is {\it fast}!
Since $\eta^\prime$ is the heaviest and ``stickiest" (glue rich) 
of the lowest lying mesons, it would have been last on the list of
possible fast, leading particles in $B$ decay searches.
There is one thing unique to $\etap$, however,
namely its connection to the gluon anomaly.
$\eta$-$\etap$ mixing is said to be related to the axial U$(1)$ problem,
and the symmetry is broken by the $G\tilde G$ gluon anomaly.
Indeed, in the chiral limit of $m_q\rightarrow 0$ 
(assuming $N_F = 3$ of light flavors),
one has 
$\langle 0\vert \partial_\mu J^0_{\mu 5} \vert \etap\rangle
= \langle 0\vert (2N_F\alpha_s/4\pi){\rm tr}(G\tilde G)\vert\etap\rangle$,
and it is this large, {\it topological} glue content of
$\etap$ that makes it so heavy.
So, is the $\etap$ production linked to $\bsg$?

\subsection{$\etap$-$g$-$g$ Coupling and Need for $\bsg \sim 10\%$}

Atwood and Soni \cite{AS} (AS) have indeed made such a connection,
linking $\bsg^*$ to inclusive $\etap$ via the $\etap$-$g$-$g$ gluon anomaly.
Defining the phenomenological coupling
$H(q^2,k^2,m_{\eta^\prime}^2)\, \varepsilon_{\mu\nu\alpha\beta}\, 
q^\mu k^\nu\varepsilon^\alpha(q)\varepsilon^\beta(k)\delta^{ab}$,
they extract $H(0,0,m_{\eta^\prime}^2) \simeq 1.8$ GeV$^{-1}$ 
from $J/\psi\to \eta^\prime\gamma$ decay.
Assuming that 
$H(q^2,0,m_{\eta^\prime}^2) \approx H(0,0,m_{\eta^\prime}^2)$ is constant,
they find that the SM $\bsg^* \to sg\etap$ could account for Eq. (3).
However, they seem to have mistaken 
$d\Gamma/dq$ for $d\Gamma/dm$, 
where $m = m_{X_s} \equiv m_{\rm recoil}$.
They hence have a false sensitivity to Fermi motion.
Furthermore, the assumption of constant $H(q^2,k^2,m_{\eta^\prime}^2)$
is definitely too strong.

The $q^2$ ranges from 0 to $m_b^2$, way beyond the QCD scale
that determines $m_{\etap}$.
We shall assume that form factor effects do not set in,
which in itself is already a big assumption.
But even then, for such a broad range of $q^2$,
one does not expect couplings to stay constant.
This is especially so since one finds that
$d\Gamma/dq$ peaks at large $q > 3$ GeV (Fig. 2(b)).

So, let us try \cite{HT} to 
understand the $\etap$-$g$-$g$ coupling better.
The $\etap$ problem
in QCD is in itself an active field of research.
The anomaly coupling comes from the Wess-Zumino term,
without assuming PCAC and soft pions,
\begin{equation}
-i\, a_g \ c_P\, \eta^\prime
\, \varepsilon_{\mu\nu\alpha\beta}\, 
\varepsilon^\mu(q) \varepsilon^\nu(k) q^\alpha k^\beta,
\end{equation}
where
$a_g(\mu^2) = {\sqrt{N_{\rm F}}\, \alpha_s(\mu^2)/ \pi f_{\eta^\prime}}
\equiv H(q^2,k^2,m_{\eta^\prime}^2)$ of AS.
The explicit $\alpha_s$ factor strongly suggests that 
one should use running coupling.
In the case at hand, since $k^2 \to 0$ and
$q^2> m^2_{\etap}$ is the dominant scale,
we expect $\mu^2 = q^2$.
As a cross check, we find that 
$a_g(m_{\eta^\prime}^2) \simeq 1.9$ GeV$^{-1}
\simeq H(0,0,m_{\eta^\prime}^2)$ of AS,
but for larger $q^2$, this would suppress the SM $b\to sg^*$ effect,
since the strong coupling
changes by a factor of 2 from $m_{\etap} \sim 1$ GeV to $m_b$ scale.
We find \cite{HT} a factor of 1/3 suppression of
SM $\bsg^* \to sg\etap$,
hence $\bsg \approx 10\%$ is precisely what is called for.
It is interesting that both $\vert F_1^{\rm SM}\vert \sim 5$
and $\vert F_2^{\rm New}\vert \sim 2$ effects are needed.
Furthermore, 
because in the SM the $F_1$-$F_2$ interference effect is destructive,
the sign of $F_2^{\rm New}$ should be opposite to that in the SM,
which is precisely what is found in the SUSY example of
$\bsg \sim 10\%$ \cite{CGG}!

\begin{figure}[h]
\includegraphics{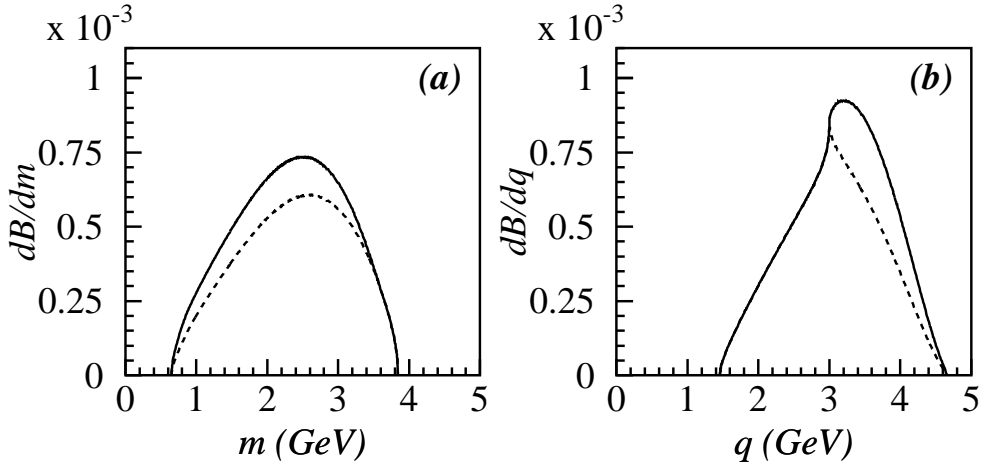}
\vskip 5.8cm
\caption{
(a) $d{\cal B}/dm$ and (b) $d{\cal B}/dq$ for $b\to \eta^\prime sg$ (solid) 
and $\bar b\to \eta^\prime \bar sg$
(dashed).
The $c\bar c$ threshold is evident in (b),
while the CP asymmetry  is due to new phase $\sigma$ in $F_2$.}
\end{figure}
%\vskip -.5cm

\subsection{BONUS: Potential for New $a_{\rm CP} \ \sim 10\%$}

For inclusive $b\to s$ transitions, it is known that \cite{GHp} 
$a_{\rm CP} \ltap 1\%$, due to the smallness of  
$v_u \equiv V_{us}^* V_{ub}\sim \lambda^5$  and unitarity constraints.
However, because of the appearance of the $b_R$ field in Eq. (1),
the $F_2^{\rm New}$ dipole coupling 
probes new phases that are independent of CKM phase.
At the same time, the $F_1^{\rm SM}$ coupling, which is needed
also to acocunt for the rate, 
provides the necessary rescattering phase,
from the $c\bar c$ cut in Fig. 1(a).
The differential BR's $d{\cal B}/dm$ and $d{\cal B}/dq$
are shown in Fig. 2, assuming phase difference $\sigma = 90^\circ$ 
between $F_1$ and $F_2$. 
Note that, thanks to the anomaly $\etap$-$g$-$g$ coupling,
although $q^2$ is not a physical variable,
$m^2$ directly corresponds to the physical recoil mass
against $\etap$.
Furthermore, large $q^2$ (hence fast $\etap$)
is favored by the anomaly coupling!
The upshot is that
we can account for the huge branching ratios that are
{\it already observed},
while the asymmetry $a_{\rm CP} \sim 10\%$ in
$B \to \etap + K^\pm + X$. % (or, sign in $\pi^\pm$).
In principle, CLEO could probe this asymmetry very soon.

\section{Conclusion}

The SM expectation that $b\to sq\bar q \sim 1\%$ and
$\bsg \sim 0.2\%$ is quite firm.
However, persistent $\Bsl$ and $n_C$ problems
hint at the possibility of $\bsg \sim 10\%$ from new physics.
The recent observation of spectacularly large 
semi-inclusive $B\to \etap + X_s \sim 0.75\times 10^{-3}$
where $p_{\etap} > 2$ GeV poses an additional challenge to the SM.
It is proposed that large $\bsg$ leads to large $\Betap$ through
the gluon anomaly.
We find that, 
with running $\alpha_s$ in the $g^*$-$g$-$\etap$ coupling, 
both SM $\bsg^* \sim 1\%$ and new physics $\bsg \sim 10\%$ are needed,
to feed down to $\Betap$.
The anomaly coupling preferentially leads to fast $\etap$ mesons.
Since the new physics color dipole transition involves right-handed couplings
to the $b$ quark,
one probes a new CP violating phase that is independent of CKM.

The $\Betap$ mode is already observed at 0.1\% level.
With 10\% $a_{\rm CP}$ possible because of the interplay of SM and new physics,
perhaps CP violation could be observed before 1999,
the year that B Factories turn on.


\begin{references}
\bibitem{Poling}CLEO numbers are from the talk by Poling, R., these proceedings.
\end{references}
\end{document}